\documentclass[journal,onecolumn]{IEEEtran}
\usepackage{authblk}

\usepackage{cite}

\usepackage{url}

\ifCLASSINFOpdf
  \usepackage[pdftex]{graphicx}
\else
\fi

\usepackage{float}
\usepackage{amsmath}

\begin{document}
%
\title{Reverse Designing Ferroelectric Capacitors with Machine Learning-based Compact Modeling}
%
%
%

\author[1]{Diego~Ferrer}
\author[1]{Jack~Hutchins}
\author[2]{Revanth~Koduru}
\author[2]{Sumeet~Kumar~Gupta}
\author[1]{Admedullah~Aziz}
\affil[1]{Department of Electrcial Engineering and Computer Science, University of Tennessee, Knoxville}
\affil[2]{Elmore Family School of Electrical and Computer Engineering, Purdue University}

\maketitle

\begin{abstract}
Machine learning-based compact models provide a rapid and efficient approach for estimating device behavior across multiple input parameter variations. In this study, we introduce two reverse-design algorithms that utilize these compact models to identify device parameters corresponding to desired electrical characteristics. The algorithms effectively determine parameter sets, such as layer thicknesses, required to achieve specific device performance criteria. Significantly, the proposed methods are uniquely enabled by machine learning-based compact modeling; alternative computationally intensive approaches, such as phase-field modeling, would impose impractical time constraints for iterative design processes. Our comparative analysis demonstrates a substantial reduction in computation time when employing machine learning-based compact models compared to traditional phase-field methods, underscoring a clear and substantial efficiency advantage. Additionally, the accuracy and computational efficiency of both reverse-design algorithms are evaluated and compared, highlighting the practical advantages of machine learning-based compact modeling approaches.
\end{abstract}


%
\IEEEpeerreviewmaketitle

\section{Introduction}
The process of developing new electronic devices requires thorough characterization and testing to assess their performance and advantages relative to established technologies. Essential to this evaluation are detailed property measurements and rigorous circuit simulation analyses. As device complexity grows, the traditional approach of physical prototyping becomes increasingly resource-intensive, making computational simulations a vital component of modern device development.

Initial stages typically involve device-level simulations, employing advanced methods such as atomistic simulations, which capture interactions at an atomic scale, or technology computer-aided design (TCAD) simulations, rooted entirely in physics-based principles. While these methods offer highly detailed insights, they are computationally demanding and impractical for direct incorporation into extensive circuit-level evaluations involving numerous device instances.

To bridge this gap, detailed simulation models must be transformed into compact models, optimized for efficient integration into circuit simulations. Traditionally, compact model development involves constructing physics-based equations representative of device behavior and subsequently fine-tuning numerous parameters to achieve desired accuracy. Although physics-based compact models generally offer high accuracy, the extensive parameter tuning required can span months, demanding significant expertise and resources.

Furthermore, increasing device complexity leads to growth in the number of necessary model parameters. When these physics-based compact models encompass numerous parameters to maintain high accuracy, they frequently revert to computationally intensive forms, restricting their practical application in extensive circuit simulations \cite{DataDrivenCompactModels}.

An alternative approach to physics-based compact modeling is the use of data-driven models. Rather than relying on explicit physics-based equations, these data-driven models leverage experimental or simulation-generated data and apply machine learning techniques to predict device behavior \cite{dnnmodelexplanation}. Data-driven modeling provides significant advantages, notably in reducing the complexity and computational expense associated with traditional physics-based methods \cite{dnnmodelbasedadvantages}. These attributes enable innovative applications and novel approaches to device design and optimization, which constitute the primary focus of this study.

The utilization of efficient machine learning-based compact models significantly accelerates the prediction of device behavior, making rapid exploration of device parameters practical. This capability becomes increasingly valuable considering that minor alterations in a device's physical structure, such as variations in layer thicknesses, can profoundly influence its overall behavior. Traditional compact modeling methods often become prohibitively time-consuming when precise behavioral specifications must be met, particularly when iterative parameter adjustments are required. Therefore, the intersection of stringent behavioral requirements and the necessity for swift parameter determination represents a prime opportunity for the application of the methods proposed in this study.

The key contributions of this work are as follows:

\begin{itemize}
\item Development of a machine learning-based compact model for accurately predicting the behavior of a novel electronic device across various device parameter configurations.
\item Introduction of two reverse-design algorithms capable of efficiently identifying optimal device parameters necessary to achieve specified electrical characteristics.
\end{itemize}

\section{Related Work}
Extensive research has been undertaken in the domain of machine learning-based compact modeling, particularly for emerging device architectures inadequately represented by conventional models such as the Berkeley Short-channel IGFET Model (BSIM). For example, machine learning approaches have been employed to model devices like Reconfigurable Field-Effect Transistors (RFETs), which exhibit complex behavior necessitating alternative modeling methodologies \cite{reconfigFETCompactModels}. Additionally, prior studies have introduced Artificial Neural Network (ANN)-based compact models tailored explicitly for novel architectures, such as Nanosheet Field-Effect Transistors (NSFETs) \cite{ANNCMForNovelNSFETDevice}. While this work similarly leverages ANN methodologies for compact modeling, it distinguishes itself by targeting a different device category and integrating the developed compact model into a reverse-design framework.

As the field of machine learning-based compact modeling expands, increasing attention has been directed toward optimizing neural network design methodologies. Recent studies have investigated advanced neural network architectures—including Mixture Density Networks, Graph Neural Networks, and Physics-Inspired Neural Networks—to enhance modeling accuracy and efficiency \cite{NovelMDNModelforHeaterCrytrons, NovelGNNModelforFinFETs, NovelPhysicsInpiredModelforThin-TFET, PriorPhysicsKnowledgeBasedFinFETCompactModels, PhysicsInspiredRRAMCompactModels}. Concurrently, research efforts have been made toward establishing generalized workflows applicable across diverse device classes, exemplified by studies focused on modeling parasitic capacitances and stochastic device behaviors \cite{GeneralizedWorkflowForMLCMForMultiStateDevices, InterconnectCompactModels}. However, this study differentiates itself by placing less emphasis on specific neural network architectures or generalized workflows. Instead, it prioritizes the practical application of a developed compact model within a reverse-design algorithm, aiming to determine optimal device parameters efficiently.

Furthermore, some existing research highlights broader applications of machine learning-based compact models beyond direct device characterization. For instance, these models have been applied to enhance circuit convergence analysis in SPICE simulations, addressing challenges that traditional compact modeling techniques struggle to overcome \cite{MLCMSPICEConvergenceStudy}. While sharing commonalities in utilizing machine learning-based compact models as fundamental components, this study uniquely integrates such models within a reverse-design methodology specifically for the optimization of ferroelectric capacitor parameters.

\section{Methods}
The initial step in this study involved the development of a machine learning-based compact model utilizing a neural network framework. This approach aimed to accurately capture variations in device behavior resulting from adjustments in key physical parameters.

Following the establishment of the compact model, two reverse-design algorithms were implemented: one algorithm based on Bayesian optimization techniques and another employing gradient-based optimization methods.

To evaluate and validate these approaches, the algorithms were specifically tested using a Metal-Ferroelectric-Metal (MFM) capacitor, as illustrated in Fig. 1. The MFM capacitor represents an innovative variant of ferroelectric capacitors, which are increasingly important components in advanced devices such as ferroelectric field-effect transistors (FEFETs), ferroelectric tunnel junctions (FTJs), and ferroelectric random-access memories (FERAMs). These devices hold significant potential for applications including multi-level memory storage, in-memory computing, neuromorphic computing, steep-slope switching, and low-power electronics \cite{FEFET1, FEFET2, FEFET3}.

Although initially developed and demonstrated with the MFM capacitor, the reverse-design algorithms were intentionally generalized to facilitate applicability across a diverse range of electronic devices.

\begin{figure}
    \centering
    \includegraphics[width=0.5\linewidth]{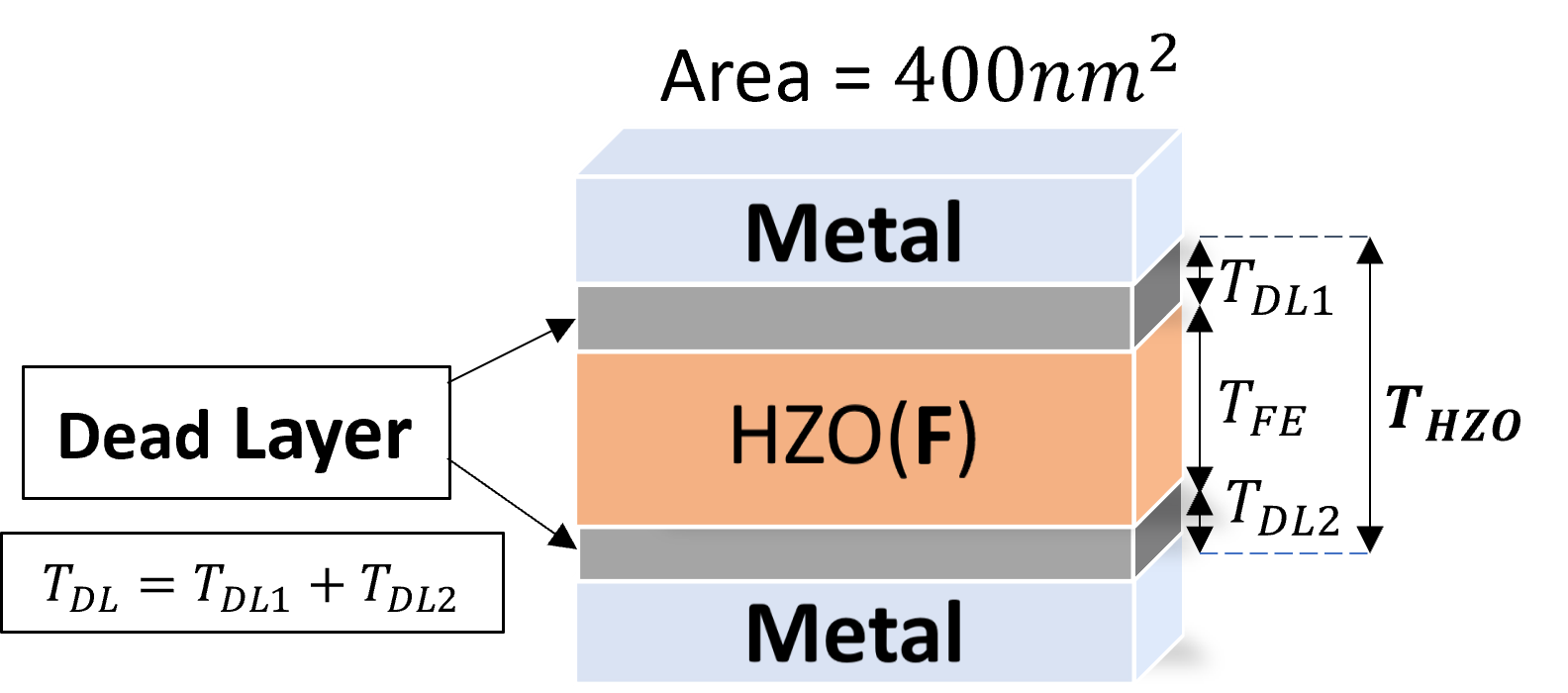}
    \caption{This figure illustrates the structural design of the Metal-Ferroelectric-Metal (MFM) capacitor, a novel approach to ferroelectric capacitor fabrication. Phase-field simulation data obtained from this device configuration were utilized as training data for the machine learning-based compact model developed in this study.}
    \label{fig:MFM_Model}
\end{figure}

\subsection{Machine Learning-Based Compact Model}
The performance characteristics of the Metal-Ferroelectric-Metal (MFM) capacitor can significantly differ depending on the thicknesses of the ferroelectric and dielectric "dead" layers. To accurately capture this variability, the machine learning-based compact model was trained using data generated from phase-field simulations covering multiple combinations of these layer thicknesses \cite{D4NR03700F}. To assess the model's predictive capability for untrained conditions, a subset of the simulation-generated data was reserved explicitly for validation and testing purposes. From this subset, a target electrical behavior was selected to serve as the optimization objective for the developed reverse-design algorithms.

\begin{figure}
    \centering
    \includegraphics[width=0.75\linewidth]{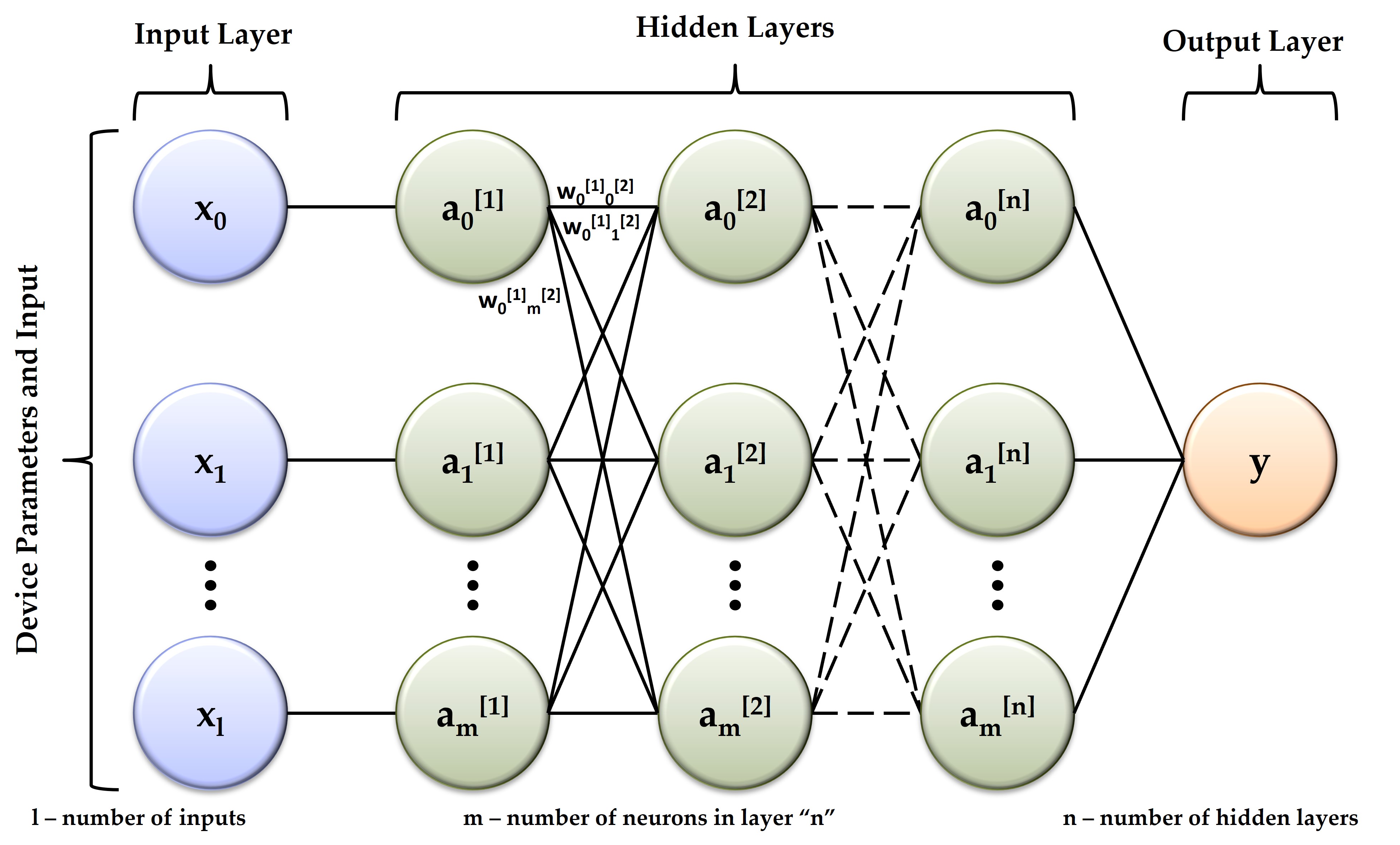}
    \caption{This figure illustrates the general architecture of the neural network employed within the reverse-design algorithm. The notation introduced here will be referenced throughout subsequent sections of this study.}
    \label{fig:NeuralNetwork}
\end{figure}

Multiple neural network architectures were evaluated, varying in the number of neurons, layers, and neuron distribution per layer. Given that the neural network would be integrated into an iterative optimization framework, a balance was established between prediction accuracy and computational efficiency. The optimal neural network configuration identified consisted of four hidden layers, each comprising 512 neurons. To further enhance accuracy, several training techniques were explored, including the introduction of Gaussian noise to create additional data points \cite{GeneralizedWorkflowForMLCMForMultiStateDevices}. The neural network was implemented using Python, specifically leveraging the Keras and TensorFlow libraries, which provided the necessary functionalities for constructing layers, training the model, and generating predictions \cite{keras, tensorflow}.

\begin{figure}
    \centering
    \includegraphics[width=0.75\linewidth]{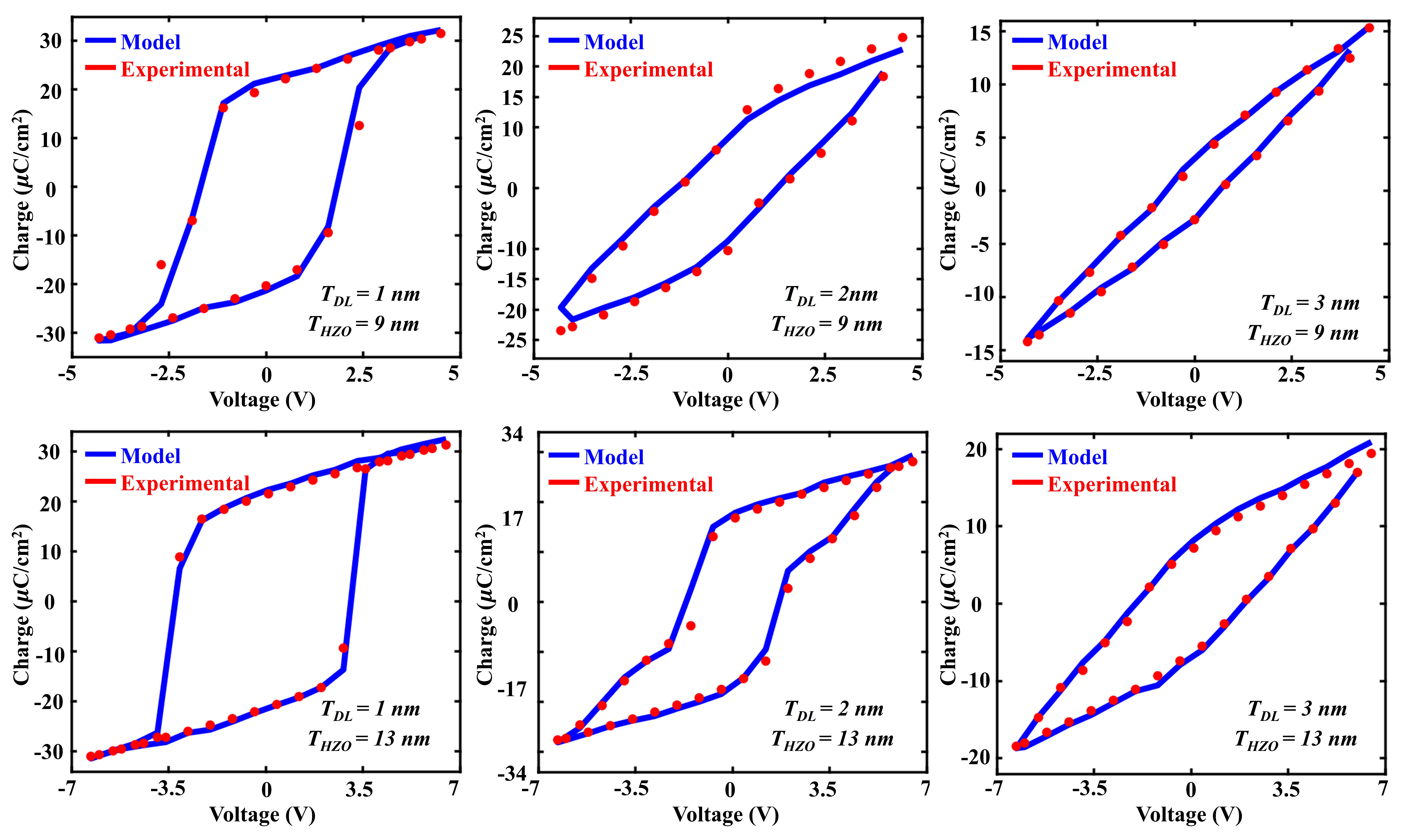}
    \caption{This figure illustrates the accuracy of the neural network model by comparing its predicted values to phase-field simulation data for the MFM capacitor. The model achieved a coefficient of determination (R²) of 0.991 and a root mean squared error (RMSE) of 1.967, indicating high predictive accuracy.}
    \label{fig:ModelTestAccuracy}
\end{figure}

\subsection{Reverse Design Algorithm}
Following the development of the machine learning-based compact model, which accurately predicts device behavior across various parameter combinations, we integrated the model into multi-parameter reverse-design algorithms. The objective of these algorithms is to identify the specific set of device parameters required to achieve a targeted electrical behavior. This is accomplished by initially generating predictions of optimal device parameters that could yield the desired device characteristics. Subsequently, the compact model assesses the predicted parameters to estimate device performance. The resulting predicted behavior is then compared against the target behavior, with the discrepancy between these behaviors quantified as a loss metric guiding subsequent parameter predictions. The iterative prediction and evaluation cycle continues until the algorithm either reaches a satisfactory minimum loss or converges to a stable set of device parameters. To rigorously evaluate algorithm performance, we conducted 50 independent trials using randomized initial conditions and target behaviors, allowing for the calculation of average algorithm efficacy and stability.

\subsubsection{Bayesian-based Method}
The Bayesian-based reverse-design method was developed by integrating Bayesian optimization principles with a machine learning-based compact model. Implementation utilized a Python-based Bayesian optimization library to facilitate efficient algorithm execution \cite{gardner2014bayesian}.

Parameter bounds for the optimization were defined using the minimum and maximum values derived from the training dataset. The algorithm began by evaluating an initial set of randomly generated device parameters drawn from a Gaussian process prior. Subsequently, the machine learning-based compact model calculated the corresponding electrical characteristics, which were evaluated against target characteristics using the coefficient of determination (R²) as the performance metric.

In each iteration, new parameter sets were sampled and evaluated, with outcomes progressively incorporated into the posterior distribution. This iterative process systematically refined parameter estimations, guiding the algorithm toward optimal device parameters. For the purposes of this research, the algorithm iterations continued until either the predicted electrical characteristics reached an R² accuracy of at least 0.97 or the process reached the maximum limit of 1000 iterations.

\subsubsection{Gradient-based Method}
The gradient-based reverse-design method employs principles derived from the backpropagation algorithm commonly used in neural network training. Typically, backpropagation adjusts network weights by calculating the gradient of a defined loss function and updating the weights accordingly:

\[
W_f = W_i - \alpha\Delta L
\]

A frequently utilized loss function is the Mean Squared Error (MSE), also referred to as L2 Loss, expressed as:

\[
L = \frac{1}{n}\sum_{i=1}^{n}(y^{(i)}-\hat{y}^{(i)})^2
\]

This equation computes the difference between the neural network’s predicted outputs and the target values across \( n \) data points:

\[
y^{(i)} = a_0^{[m]}w_0^{[m]}\vphantom{w}_y + \dots + a_n^{[m]}w_n^{[m]}\vphantom{w}_y
\]

Traditionally, the gradient of this loss is calculated with respect to the neural network's weights to enhance accuracy. However, assuming the neural network model has already attained satisfactory accuracy, adjustments can instead be directed toward the device parameters themselves.

To facilitate reverse design of device parameters \( \theta \), the gradient of the loss function relative to these parameters must be determined. First, the error at the output layer is defined as:

\[
\delta^{(L)} = \frac{\partial \mathcal{L}}{\partial y} \odot \sigma^{(L)'}\Bigl(z^{(L)}\Bigr),
\]
where \( \odot \) denotes the Hadamard (element-wise) product, and \( \sigma^{(L)'} \) represents the derivative of the activation function at layer \( L \).
For preceding layers \( l = L-1, L-2, \dots, 1 \), the error is computed recursively as:

\[
\delta^{(l)} = \left(W^{(l+1)T}\,\delta^{(l+1)}\right) \odot \sigma^{(l)'}\Bigl(z^{(l)}\Bigr).
\]

Considering the device parameters \( \theta \) are components of the neural network input \( a^{(0)} \), the error propagated back to the input layer is calculated by:

\[
\delta^{(0)} = W^{(1)T}\,\delta^{(1)}.
\]

Since \( a^{(0)} = \begin{bmatrix} V \\ \theta \end{bmatrix} \), extracting the components corresponding to the device parameters yields:

\[
\frac{\partial L}{\partial \theta} = \Bigl[\delta^{(0)}\Bigr]_{\theta},
\]
where the notation \(\Bigl[\delta^{(0)}\Bigr]_{\theta}\) denotes the specific components of \( \delta^{(0)} \) related to \( \theta \).

The general procedure for this algorithm initiates with a set of initial device parameters, input into the pre-trained neural network, which predicts the resultant device behavior. The predicted behavior is compared to the target behavior using the MSE loss function, facilitating the computation of the gradient of the loss. Subsequently, the product of this gradient and a predefined learning rate is subtracted from the current parameters, steering them iteratively toward optimal values. This iterative adjustment process continues until the predicted device parameters yield electrical characteristics with an \( R^2 \) accuracy of at least 0.97 relative to the target behavior or until a maximum of 1000 iterations is reached.

\begin{figure}
    \centering
    \includegraphics[width=0.8\linewidth]{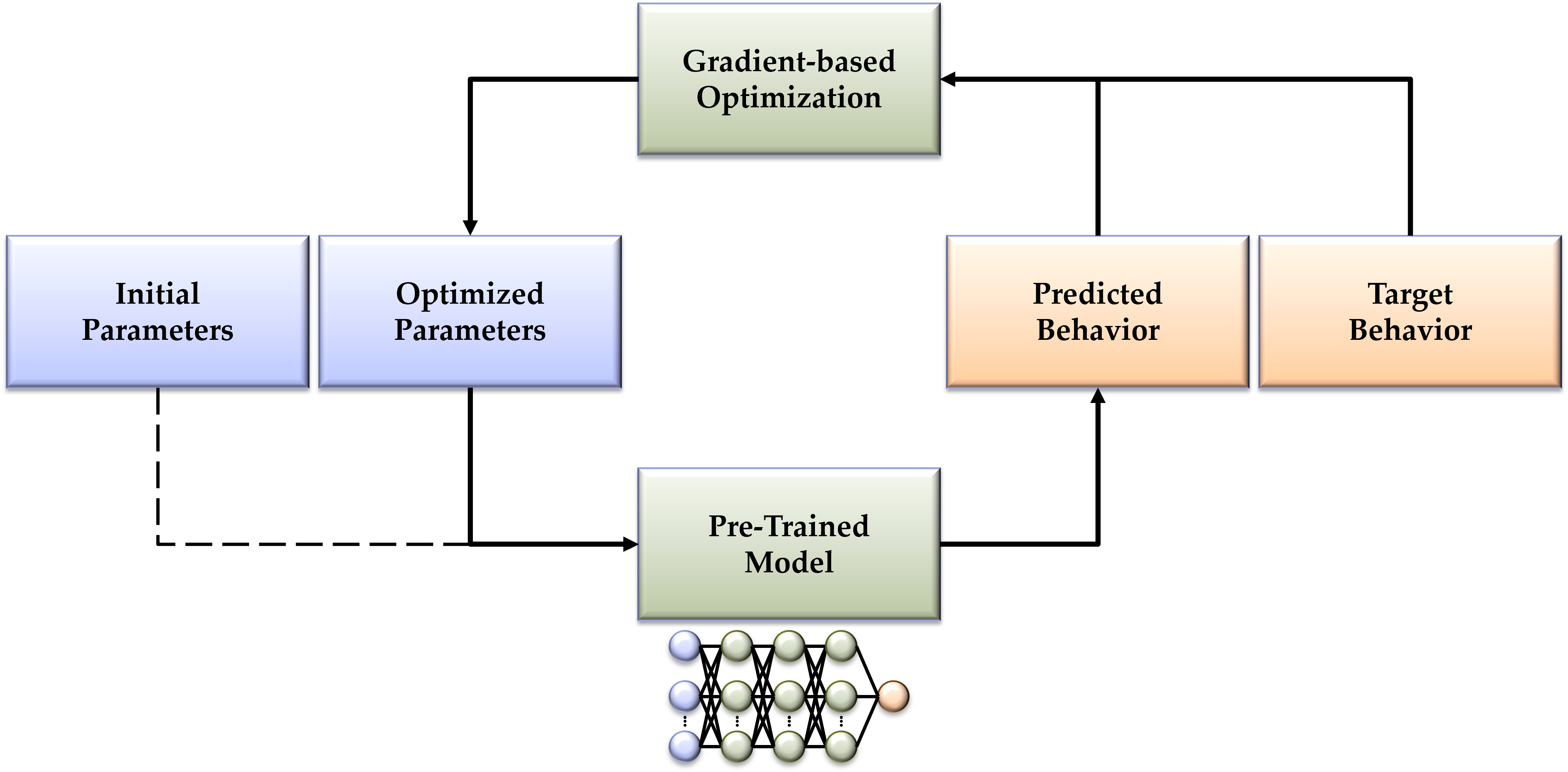}
    \caption{The gradient-based method begins by randomly selecting an initial set of design parameters within defined bounds. These parameters are input into a neural network trained on simulation or experimental data generated for different parameter configurations, which predicts the device’s behavior. The predicted behavior is then compared with a target behavior, and the discrepancy is used to compute the gradient of the loss function with respect to the parameters. The parameters are updated iteratively using this gradient, and the cycle of prediction and adjustment continues until a convergence criteria is met—either the predicted behavior sufficiently matches the target behavior or a maximum number of iterations is reached.}
    \label{fig:GradOptimization}
\end{figure}

\section{Results}
The primary focus of this analysis includes evaluating the accuracy with which the reverse-design algorithms determine device parameters capable of achieving specified target behaviors, as well as measuring the computational time required for the algorithms to converge to these optimal parameters. Demonstrating the accuracy underscores the reliability and effectiveness of the algorithms in identifying suitable parameter sets. Additionally, the comparative analysis of computational time highlights the substantial advantage of employing machine learning-based compact models over traditional, computationally intensive phase-field simulations.

\subsection{Algorithm Accuracy}
In conducting the reverse-design algorithms, results were systematically collected using various initial conditions and target values. Upon selection of these target values, corresponding data were removed from the neural network training set to prevent data leakage, ensuring the integrity of the evaluation. Each reverse-design algorithm provided solutions based on either achieving a predetermined minimum accuracy or reaching a maximum of 1000 iterations.

Representative results from the Bayesian-based algorithm are illustrated in Fig. 5, while Fig. 6 depicts outcomes from the gradient-based method. A visual comparison indicates that, under identical conditions, the gradient-based algorithm consistently yielded parameters more closely aligned with target behaviors. Specifically, results shown in Fig. 5(a) exhibit notably poorer performance compared to those in Fig. 6(a), as evidenced by the coefficient of determination (R²) values. This discrepancy highlights a notable difficulty for the Bayesian approach in improving beyond a certain accuracy threshold, affirming the gradient-based method's superior consistency.

Additionally, Fig. 6(a) prominently displays step-like switching behavior in the phase-field simulation data. This phenomenon is attributed to the small dimensions of the simulated device, resulting in a limited number of grains and domains and subsequently causing discrete, step-like switching characteristics \cite{StepLikeSwitching}. Smoother switching behavior would potentially be observed by incrementally refining voltage steps in simulations or performing multiple cycles and simulating variations across devices. However, these adjustments would necessitate extensive computational resources. Given that the performance of the machine learning-based compact model is largely insensitive to these specific variations, additional intensive simulations were considered unnecessary for the scope of this study.

\begin{figure}
    \centering
    \includegraphics[width=0.75\linewidth]{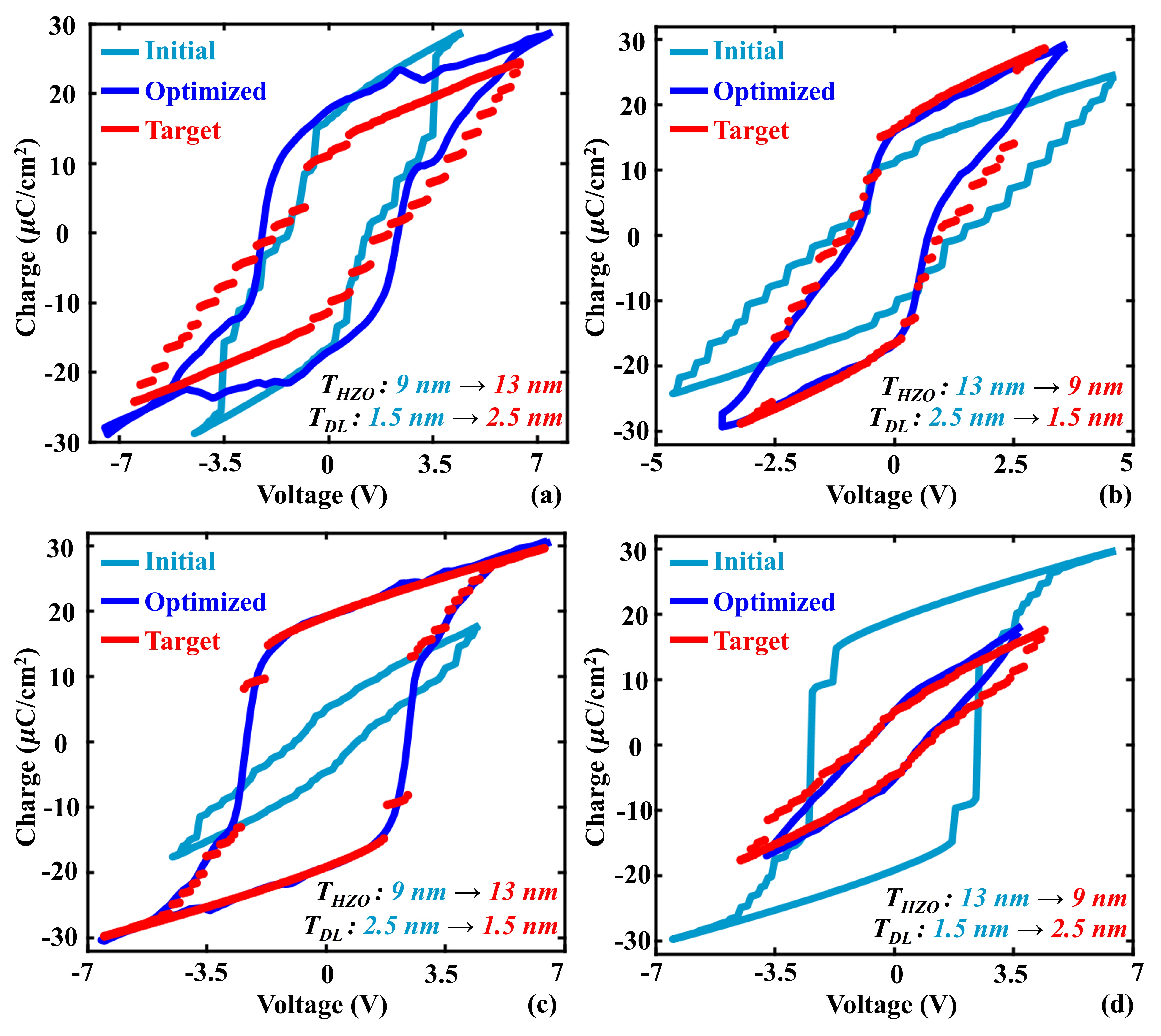}
    \caption{This figure shows the results of the Bayesian optimization method for four different runs. The lighter blue plots show the graph created by the model for the initial set of parameters. The algorithm begins with these parameters and optimizes towards the red target values. The values of the initial and target values for each device parameter are shown in the bottom right corner of each sub-figure. The ending accuracy of each run are as follows: (a) $R^2$: 0.931, RMSE: 5.168; (b) $R^2$: 0.978, RMSE: 2.791; (c) $R^2$: 0.996, RMSE: 1.433; (d) $R^2$: 0.967, RMSE: 2.285}
    \label{fig:BayesianResults}
\end{figure}

\begin{figure}
    \centering
    \includegraphics[width=0.75\linewidth]{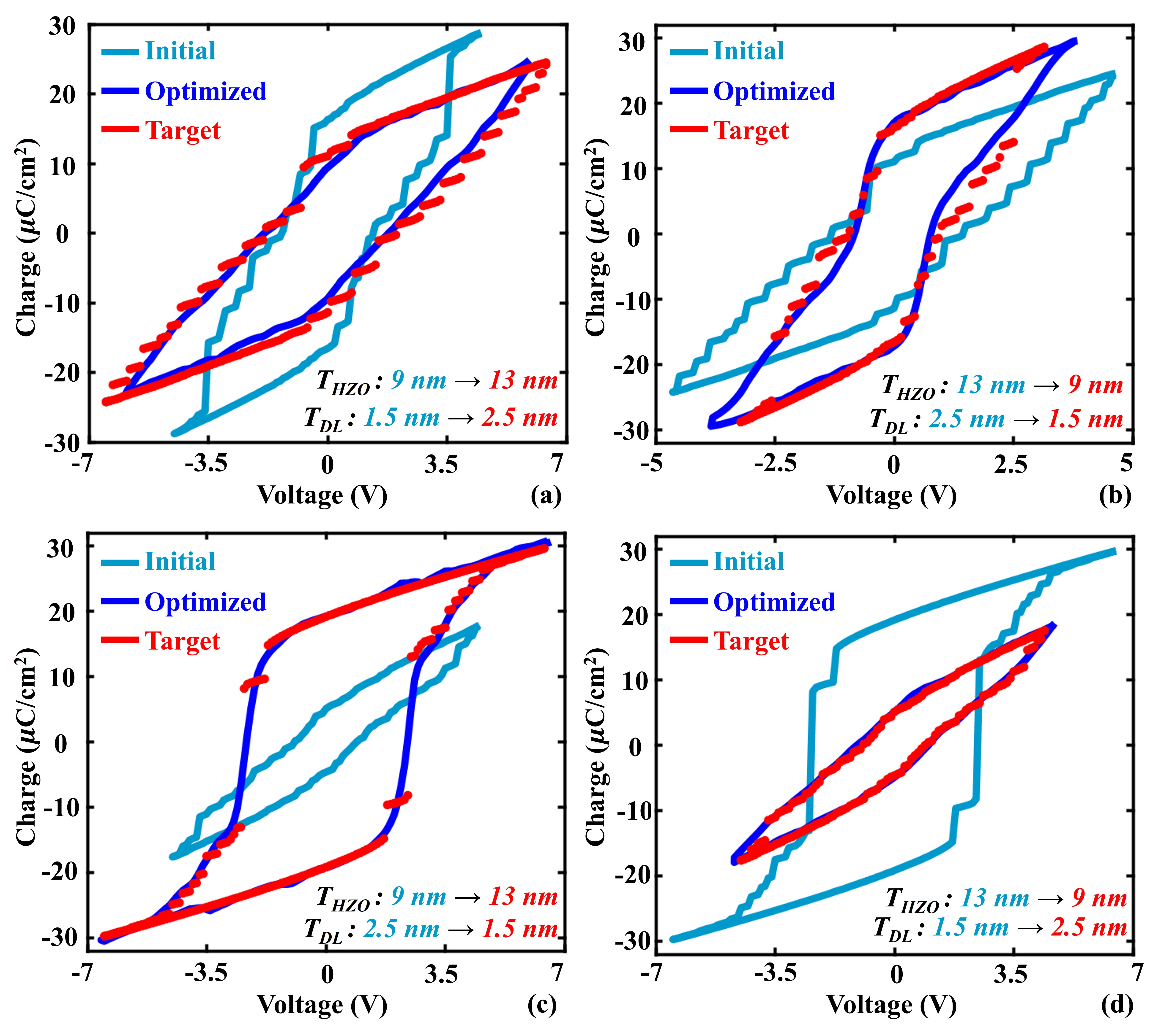}
    \caption{This figure shows the results of the Gradient optimization method for four different runs. The lighter blue plots show the graph created by the model for the initial set of parameters. The algorithm begins with these parameters and optimizes towards the red target values. The values of the initial and target values for each device parameter are shown in the bottom right corner of each sub-figure. The ending accuracy of each run are as follows: (a) $R^2$: 0.990, RMSE: 1.584; (b) $R^2$: 0.978, RMSE: 2.817; (c) $R^2$: 0.996, RMSE: 1.460; (d) $R^2$: 0.997, RMSE: 0.584}
    \label{fig:GradientResults}
\end{figure}

\subsection{Time Comparison}
A critical aspect of this study involves demonstrating the efficacy and computational efficiency of the proposed reverse-design algorithms, particularly highlighting their significant time-saving advantages compared to traditional methods. Consequently, a comparative analysis of total computational time was conducted, contrasting the machine learning-based compact model with conventional phase-field simulation approaches.

While theoretically feasible, directly performing the reverse-design process with phase-field simulations would be prohibitively time-consuming. Therefore, this study estimates the computational time by calculating the number of cycles required for the gradient-based algorithm and combining this with estimations of the time necessary for phase-field simulations to produce device behavior predictions. Phase-field simulations were conducted using MATLAB on a system equipped with an AMD EPYC 7542 32-Core processor operating at 2.9 GHz. The machine learning-based compact model and optimization algorithms were executed on a desktop computer with an 11th Gen Intel(R) Core(TM) i7-11700 CPU operating at 2.5 GHz.

The computational time required for generating voltage sweeps via phase-field simulations varied significantly depending on device structure. For instance, simulating a device with a dead layer (DL) thickness of 0.5 nm and a ferrous layer (FL) thickness of 2 nm required approximately 942 seconds, whereas a device with the same DL but an FL thickness of 14 nm necessitated roughly 63,350 seconds. In contrast, the machine learning-based compact model completed the 0.5 nm DL and 2 nm FL simulation in only 0.067 seconds and the 0.5 nm DL and 14 nm FL simulation in merely 0.053 seconds. Furthermore, the number of data points generated by these simulations depended on device geometry, with thicker ferrous layers producing broader hysteresis loops, thus enabling more extensive voltage sweeps at 0.1 V increments. For example, the 0.5 nm DL and 4 nm FL device generated 102 data points, while the 0.5 nm DL and 14 nm FL device generated 302 data points.

Considering the extensive computational time required—ranging from approximately 16 minutes in optimal scenarios to over 17 hours in worst-case scenarios—comparisons were conducted using the best, worst, and average case scenarios. Additionally, due to the absence of accessible gradients from the phase-field model, the gradient-based optimization method could not be directly employed with these simulations. Thus, the Bayesian-based algorithm was utilized, which required between 1 and 1000 iterations, averaging approximately 447 iterations per execution. Consequently, employing phase-field simulations under these conditions would result in computational durations of nearly 733 days in the worst case, approximately 16 minutes in the best case, and roughly 157 days on average.


\begin{table}
    \centering
    \small
    \caption{Cycle time represents the amount of time required to generate a voltage sweep for a cycle of the Bayesian-based algorithm. The cases for this data represent the longest, shortest, and average amount of time needed to generate this data.}
    \includegraphics[width=\linewidth]{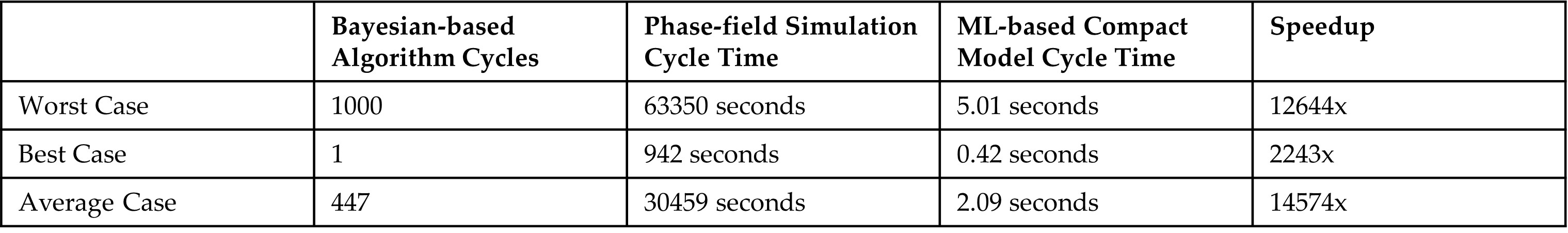}
    \label{fig:table 1}
    \vspace{-6mm}
\end{table}

An essential factor when assessing the computational time for the gradient-based algorithm is the initial overhead associated with generating the training data necessary for the machine learning-based compact model. For this research, generating sufficient training data required twenty-five phase-field simulations. The computational cost to generate this dataset ranged considerably, from approximately 6.5 hours in the best-case scenario to 440 hours in the worst-case scenario, averaging around 211 hours.

Once trained, the gradient-based algorithm required an average of 2.09 seconds per prediction cycle. With an average iteration count of 447 cycles, the total computational time amounted to approximately 934 seconds, which is minimal compared to the initial data generation effort. While it is theoretically possible under optimal conditions for direct phase-field simulations to match or outperform the machine learning-based compact model, the practical scenario demonstrates substantial advantages of the machine learning approach. Specifically, the compact model offered an average speedup of approximately 17 times and reached up to 40 times improvement in the worst-case scenario.

Additionally, once the ML-based compact model is trained, altering the target behavior only necessitates the computational time for the prediction cycles (approximately 934 seconds), representing a significant reduction in computational resources for subsequent optimization processes.

\section{Conclusion}
This study presented two machine learning-based compact modeling methodologies for reverse-designing device parameters to achieve specified electrical characteristics, demonstrating their practical implementation on a novel ferroelectric capacitor. The analysis indicates that while the Bayesian optimization method occasionally yields rapid results, the gradient-based algorithm consistently provides superior accuracy and reliability. Consequently, the preference for the gradient-based method underscores the importance of employing machine learning-based compact models, as direct application of gradient-based techniques on computationally intensive phase-field simulations is impractical.

Furthermore, the results underscore the substantial computational advantages afforded by machine learning-based compact models compared to conventional phase-field simulations, achieving an average speedup of approximately 14,000 times in simulation cycle time. Although demonstrated with a ferroelectric capacitor, the methodologies described in this work can be generalized and adapted to optimize parameters for a wide variety of electronic devices.

\ifCLASSOPTIONcaptionsoff
  \newpage
\fi

\bibliographystyle{plain}
\bibliography{bibtex/bib/references}
\end{document}